\documentclass[useAMS,usenatbib]{mn2e}
\usepackage{graphicx}
\usepackage{float}
\usepackage[usenames,dvipsnames]{xcolor}

\title[Iron K Lags in PG~1244+026]{The curious time lags of PG~1244+026: Discovery of the iron K reverberation lag}
\author[Kara et al.]{E. Kara$^{1}$\thanks{E-mail:
ekara@ast.cam.ac.uk}, E. M. Cackett$^{2}$, A. C. Fabian$^{1}$, C. Reynolds$^{3}$ and P. Uttley$^{4}$\\
$^{1}$Institute of Astronomy, Madingley Rd, Cambridge CB3 0HA\\
$^{2}$Department of Physics and Astronomy, Wayne State University, Detroit, MI 48201, USA\\
$^{3}$Department of Astronomy, University of Maryland, College Park, MD 20742, USA\\
$^{4}$Astronmical Institute `Anton Pannekoek', University of Amsterdam, Postbus 94249, 1090 GE Amsterdam, the Netherlands\\}
\begin{document}

\date{\today}

\pagerange{\pageref{firstpage}--\pageref{lastpage}} \pubyear{2013}

\maketitle

\label{firstpage}

\begin{abstract}
High-frequency iron K reverberation lags, where the red wing of the line responds before the line centroid, are a robust signature of relativistic reflection off the inner accretion disc.  
In this letter, we report the discovery of the Fe K lag in PG~1244+026 from $\sim 120$~ks of data (1 orbit of the {\em XMM-Newton} telescope).  The amplitude of the lag with respect to the continuum is $1000$~s at a frequency of $\sim 10^{-4}$~Hz. We also find a possible frequency-dependence of the line: as we probe higher frequencies (i.e. shorter timescales from a smaller emitting region) the Fe K lag peaks at the red wing of the line, while at lower frequencies (from a larger emitting region) we see the dominant reflection lag from the rest frame line centroid.  The mean energy spectrum shows a strong soft excess, though interestingly, there is no indication of a soft lag. Given that this source has radio emission and it has little reported correlated variability between the soft excess and the hard band, we explore one possible explanation in which the soft excess in this source is dominated by the steep power-law like emission from a jet, and that a corona (or base of the jet) irradiates the inner accretion disc, creating the blurred reflection features evident in the spectrum and the lag. 
General Relativistic ray-tracing models fit the Fe~K lag well, with the best-fit giving a compact X-ray source at a height of $5 r_{\mathrm{g}}$ and a black hole mass of $1.3 \times 10^{7} M_{\sun}$.
\end{abstract}


\section{Introduction}
\label{intro}

Recent timing analysis of AGN has shown clear evidence for time delays at the energy of the broad Fe K emission line.  This is a model-independent result that shows that at high frequencies, the line centroid energy of the Fe K line is delayed with respect to red wing \citep{zoghbi12,kara13a,kara13b,zoghbi13,kara13c}.  

The natural interpretation for this high-frequency Fe~K lag is that it is a reverberation lag associated with reflection off the inner accretion disc \citep{guilbert88,lightman88,ross05}.  In this scenario, the X-ray emitting source irradiates the inner accretion disc causing the fluorescence of Fe K and other emission lines.  Reverberation lags are thus expected between the direct continuum source and the reprocessed emission off the disc.  Due to the proximity to the central black hole, these emission lines are relativistically broadened \citep{fabian89}.  The red wing of the line is the most gravitationally redshifted emission, and therefore originates from reflection at the smallest radii, while the rest frame line centroid of the line originates from further out on the disc.  This causes the red wing of the line to respond before the rest frame.

\begin{figure}
\begin{center}
\includegraphics[width=7.5cm]{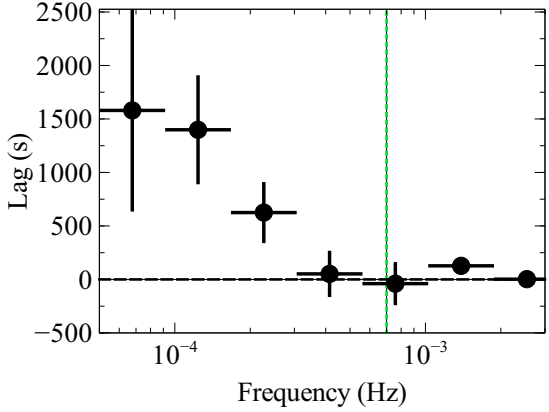}
\caption{The lag-frequency spectrum between 0.3--3~keV and 4--6.5~keV.  The green dotted line shows where the 4--6.5~keV periodogram becomes dominated by Poisson noise.  A positive lag indicates the hard band lags the soft band. The iron-line dominated hard band is seen to lag behind the continuum at all frequencies below the frequency where Poisson noise begins to dominate (and above, though we do not consider the Poisson dominated regime).}
\label{lagfreq}
\end{center}
\end{figure}

In this letter, we report the discovery of a broad Fe~K lag in the highly variable Narrow-line Seyfert I galaxy, PG~1244+026 ($z=0.0482$).  The source was originally observed with {\em XMM-Newton} in 2001 for 8~ks (PI: A.C. Fabian), and was reported to have a strong soft excess and highly ionised Fe K emission line \citep{porquet04,crummy06}.  The source was observed again in 2011 for 123~ks (PI: C. Jin) in a similar flux state \citep{jin12}, and was recently analysed in more detail for its spectral-timing properties \citep{jin13}.  In this work, the time-integrated spectrum was degenerate to both a Comptonisation continuum + neutral reflection model and a relativistically blurred reflection model.  They also report that at high frequencies (above $\sim 2 \times 10^{-4}$~Hz), the soft excess and continuum are highly variable, but the soft excess is not strongly correlated to the 4--10~keV emission on these time scales.  We expand on these spectral timing results by analysing the X-ray lag, which helps break degeneracies found from time-integrated spectral modelling alone.

This letter is organised as follows: In Section~\ref{obs}, we describe the observation and data reduction, in Section~\ref{results}, we present our results of the high-frequency lag-energy spectrum, and show its possible dependence on frequency. We then use these lag results to motivate an alternative spectral model that we fit to the mean spectrum.  Lastly, we fit the lag with General Relativistic ray tracing simulations to put independent constraints on the geometry of the system. Further discussion is presented in Section~\ref{discuss}.

\section{Observations and Data Reduction}
\label{obs}

\begin{figure}
\begin{center}
\includegraphics[width=7.5cm]{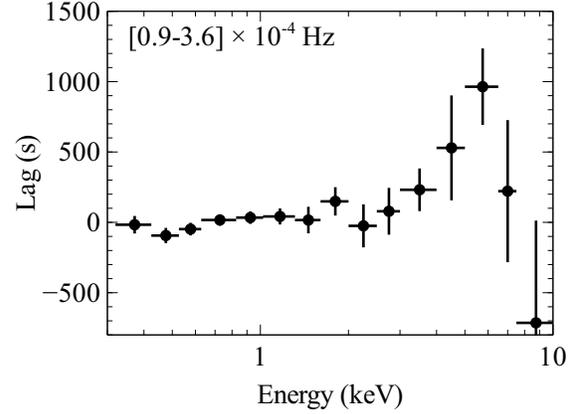}
\caption{Lag-energy spectrum at frequencies, $[0.9-3.6] \times 10^{-4}$~Hz. The lag peaks at 6.5~keV, and then turns over.  There is no apparent soft lag at energies below 1~keV.  The amplitude of the average lag at 6.5~keV is $\sim 1000$~s with respect to the 1--2~keV band that has zero lag.}
\label{totlagen}
\end{center}
\end{figure}
\begin{figure}

\begin{center}
\includegraphics[width=7.5cm]{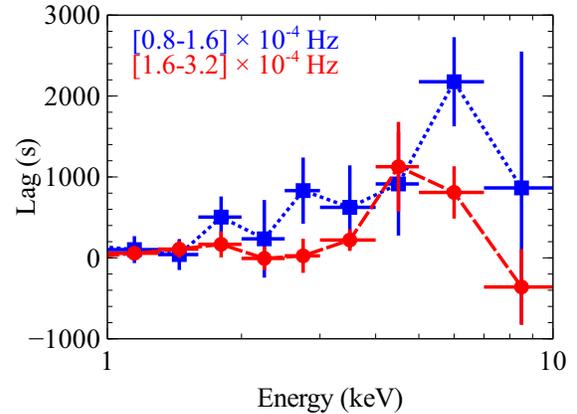}
\caption{Lag-energy spectrum at two smaller frequencies ranges: Low frequencies ($[0.8-1.6] \times 10^{-4}$~Hz; blue) and high frequencies ($[1.6-3.2] \times 10^{-4}$~Hz; red). The low frequency lag-energy spectrum peaks between 5--7 keV, while the high frequency lag peaks at 4--5~keV, and has a smaller amplitude.}
\label{freqlagen}
\end{center}
\end{figure}

PG~1244+026 was observed for $\sim$120~ks with the {\em XMM-Newton} satellite \citep{jansen01} over one full orbit on 2011 December 25 (ObsID 0675320101).  For this analysis of the reverberation lags, we focus on the high time-resolution data from the EPIC-pn camera \citep{struder01}. The observation was taken in small window mode to avoid pile-up effects. The data were reduced using the {\em XMM-Newton} Science Analysis System (SAS v.12.0.0) and the newest calibration files.  

The background was below 0.1 counts/s for the entire observation.  The source light curves were extracted from circular regions of radius 35 arcsec, which were centred on the maximum source emission. The background light curves were chosen from a circular region of the same size.  The background subtracted light curves were produced using the tool {\sc epiclccorr}.  The resulting light curve length is 123430~s with 10~s bins. %The light curve in 200~s bins is shown in Fig.~\ref{lc}.

\section{Results}
\label{results}

\subsection{Lag-Energy Spectrum}
\label{lagfreq_sec}

We calculate the lags using the standard Fourier technique detailed in \citet{nowak99}. We find the cross-spectrum for pairs of light curves, and average over frequency. At a given frequency, $f$, the argument of the cross-spectrum is the phase difference between the Fourier transforms of the two light curves.  We convert that phase lag into a frequency-dependent time lag by dividing by $2 \pi f$.  The resulting lag between the 1--3~keV and 4--6.5~keV light curves is shown in Fig.~\ref{lagfreq}.  The 4--6.5~keV band (dominated by the Fe~K line) is found to lag the continuum dominated 1--3~keV band at frequencies below $\sim 6 \times 10^{-4}$~Hz.  The 4--6.5~keV band periodogram is dominated by noise above $7 \times 10^{-4}$~Hz, and therefore, we do not look for lags above this frequency.

We follow up this hard lag by computing the lag-energy spectrum, in a manner similar to the analysis in \citet{zoghbi11}. We compute the lag between the light curve at each narrow energy bin and a reference band light curve, chosen to be the entire 0.3-10.~keV band (so as to maximize the signal-to-noise). We remove the energy bin of interest from the reference band when computing the lag at that energy bin, to ensure that the noise is not correlated.  As discussed in \citet{zoghbi13}, the chosen reference band changes where the location of zero lag is, and therefore the absolute amplitude of the lag is meaningless. Rather, it is the relative lag between energy bins that is interesting.  The larger the lag, the more delayed the emission is.  \footnote{ In this source, there is some uncorrelated component in the soft excess \citep[see Section~\ref{discuss} in this paper and ][]{jin13}, and therefore the choice of reference band slightly changes the lag at soft energies, but has no effect on the lag at the Fe~K band \citep*[See Additional Note in this paper and][ for further discussion on the lags associated with the soft excess]{alston13}.}

\begin{figure}
\begin{center}
\includegraphics[width=7.5cm]{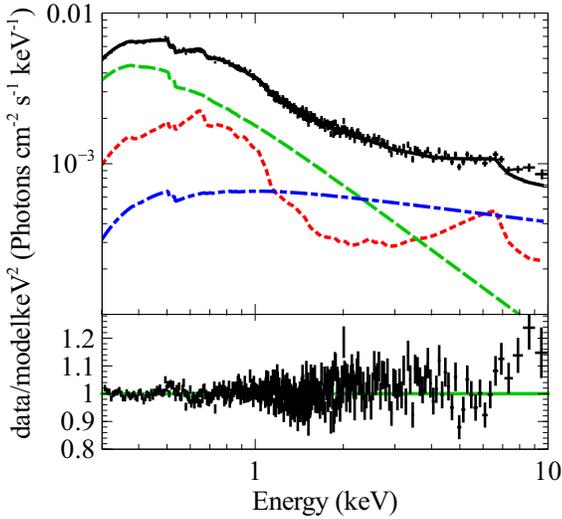}
\caption{The time-integrated energy spectrum fit with one possible ionised reflection model: {\sc relconv\_lp}$\times${\sc reflionx} (dotted red line), which is irradiated by the powerlaw continuum (blue dash-dot). The soft excess is fit by a steep powerlaw (dashed green), modelling the synchrotron tail of a relativistic jet. The bottom panel shows the data to model ratio.} 
\label{spec}
\end{center}
\end{figure}

\begin{table}
\centering
\begin{tabular}{l l l}
\hline
{\bf Component} & {\bf Parameter} & {\bf Value}\\
\hline
Galactic absorption& $N_{\mathrm{H}}(10^{22}$ $\mathrm{cm}^{-2})$ $^{\alpha}$& $0.0184$ \\
Intrinsic absorption& $N_{\mathrm{H}}(10^{22}$ $\mathrm{cm}^{-2})$ & $0.025\pm 0.0023$ \\
\hline
power law 1 & $A_{\Gamma_1} \times 10^{-3}$ &$ 1.9 \pm 0.1$\\
& $\Gamma_1$& $3.52^{+0.08}_{-0.07}$ \\
\hline
power law 2 & $A_{\Gamma_2} \times 10^{-3}$ &$ 0.8 \pm 0.1$\\
& $\Gamma_2$& $2.27 \pm 0.03$ \\
\hline
relconv\_lp & $h$ & $3.0$ $^{+0.4}_{-0.0}$\\
&$a$ &$0.97^{+0.03}_{-0.29}$ \\
&$r_{\mathrm{in}} (r_{\mathrm{g}})$& $1.6^{+0.5}_{-0.3}$ \\
&$r_{\mathrm{out}} (r_{\mathrm{g}}) $ $^{\alpha}$& 400.0 \\
&$i$ ($^{\circ}$) & $38$ $\pm 3$\\
\hline
reflionx& $A_{\mathrm{refl}} \times 10^{-8}$ &$ 8.7^{+0.5}_{-0.6}$\\
&$\xi_{\mathrm{refl}}$ (erg cm s$^{-1}$) & $498$ $^{+10}_{-30}$\\
&$Z_{\mathrm{Fe}}$ & $2.1$ $^{+0.4}_{-0.1}$ \\
\hline
$\chi^{2}/\mathrm{dof}$ & 1204/1114 = 1.08 & \\
\hline
\multicolumn{3}{l}{$^{\alpha}$ frozen}\\
\end{tabular}
\caption[Best-fit Spectral Parameters]{Best-fit spectral parameters for the time-integrated spectrum.}
\label{spec_param}
\end{table}

The lag-energy spectrum of PG~1244+026 is shown in Fig.~\ref{totlagen}.  The lag is computed in the frequency range, $[0.9-3.6] \times 10^{-4}$~Hz.  The lag-energy profile is flat below $\sim 2.5$~keV (i.e. there is no lag at soft energies).  The lag increases above 3~keV and peaks at 5--6.5~keV, beyond which, there is a sharp drop, similar in shape to the asymmetric line profiles of relativistically broadened emission lines.  The lag between 2~keV and 6~keV is about 1000~s.

We check the statistical significance of the feature at 6~keV by fitting a constant plus Gaussian model to the lag-energy spectrum from 0.3-10~keV.  We compare this model to a simple constant lag, without the additional Gaussian.  Comparing these two nested models yields an $F$-statistic of 20.4, meaning the Gaussian model is preferred with 99.91\% confidence, or over $3 \sigma$.  Interestingly, the significance of the additional Gaussian in 120~ks of PG~1244+026 is more that of Ark~564, from a 500~ks observation.

\subsubsection{Frequency-resolved Analysis}

Similar to \citet{zoghbi12} for NGC~4151, we perform a frequency-resolved analysis of the lag-energy spectrum. In that work, in addition to finding the first Fe K lag, they found that at lower frequencies, the peak of the lag-energy spectrum was at 6.5~keV, while at higher frequencies, the lag-energy spectrum peaked at $\sim 4$~keV.  In that work, the high-frequency emission was associated with emission from the smallest emitting region, which irradiated the smallest radii of the accretion disc, where the gravitational effects are strongest. Lower frequencies (longer variability timescales) probe reflection from larger radii, and thus the emission peaks at the rest frame line centroid of the broad Fe K line.  

A similar analysis of PG~1244+026 (Fig.~\ref{freqlagen}) shows similar behaviour, though the statistics in this 120~ks observation are low.  The lower frequency lag-energy spectrum in blue is from $[0.8-1.6] \times 10^{-4}$~Hz.  The amplitude of the lag between 1~keV and the peak of the emission 6~keV is $\sim$2000~s (with respect to the 1--2~keV band at zero lag).  The higher frequency lag-energy spectrum in red is from $[1.6-3.2] \times 10^{-4}$~Hz. The peak of the lag is at 4--5~keV, with an amplitude around 1000~s (with respect to the 1--2~keV band).

\begin{figure}
\begin{center}
\includegraphics[width=7.5cm]{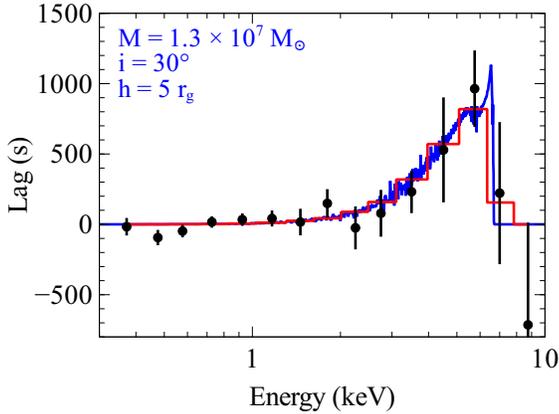}
\caption{The lag-energy spectrum of Fig.~\ref{totlagen} fit with a general relativistic model of the Fe~K lag for a source $5r_{\mathrm{g}}$ above the disc, at an inclination of $30\deg$ and a black hole spin of $a=0.998$.  The mass and dilution factor are free parameters in the fit, and are computed to be $M=1.3\times 10^{7} \mathrm{M}_{\odot}$ and $R=3.58$, respectively. The blue line indicates the model fit at high resolution, while the red shows the model binned to the resolution of the data.}
\label{lagen_mo}
\end{center}
\end{figure}

\subsection{Time-integrated Energy Spectrum}
\label{spec_sec}

In light of this new discovery of the broad Fe~K lag in PG~1244+026, we re-examine the time-integrated spectral modelling, as first described in \citet{jin13}.  In that work, the mean spectrum was not sufficient to break degeneracies between a relativistic reflection model and an additional soft X-ray continuum component. 

We propose now a new model (Fig.~\ref{spec}) to describe the time-integrated spectrum, which is consistent with the spectral-timing results of \citet{jin13} and consistent with our new result on the broad Fe~K reverberation lag.  We reiterate that this is not a unique solution, but does explain why there is a strong soft excess, but no soft lag.  

As the Fe~K lag is interpreted as reverberation between the continuum emission and the blurred emission off a highly ionised accretion disc, we use a relativistic reflection component to fit the Fe~K line (red dotted line). The model is {\sc relconv\_lp}$\times${\sc reflionx} \citep{dauser13,ross05}. While the reflection continuum contributes to the soft excess, it is not the dominant component. Below $\sim$2~keV, the spectrum is dominated by a steep power-law component (green dashed line), describing the emission from a jet component that is largely uncorrelated to the emission from small radii.  The corona (or base of the jet) irradiates the accretion disc. That component is modelled as a harder powerlaw (blue dash-dot line).  This model yields a good fit, with $\chi^2/{\mathrm{dof}}=1204/1114=1.08$. The best-fit parameters are shown in Table~\ref{spec_param}.  The parameters are quoted to 90\% confidence.

We find a source height $h<10 r_{\mathrm{g}}$, suggesting that the innermost radii are well illuminated \citep{fabian13}, and therefore the finding of a small inner radius is robust.  In this model the inner radius is assumed to be the inner most stable circular orbit, and therefore the parameters $r_{\mathrm{in}}$ and spin $a$ are not independent.

An additional narrow Fe~K line does not improve the fit. The equivalent width of such a line is $\sim 9$~eV, suggesting that there is little irradiation of cold gas at large radii.

\subsection{Fitting the lag-energy spectrum}
\label{lagmo_sec}

Finally, we fit the lag-energy spectrum of Fig.~\ref{totlagen} with the General Relativistic ray-tracing transfer functions from \citet{reynolds99}, similar to the analysis of \citet{cackett13}.  The profile of the Fe~K lag is dependent on the frequency of the Fe~K lag, the black hole mass and spin, the inclination of the disc, the height of the source above the corona, and the dilution factor. The dilution factor is the scaling due to the relative contribution of the variable reflection and variable power-law components in the chosen energy bands.    

For a lag at the frequency range specified in Fig.~\ref{totlagen}, we find for a maximally spinning black hole at a disc inclination, $i=30^{\circ}$ and a point source height, $h=5 r_{\mathrm{g}}$ above the disc, the best fit mass is $1.3 \times 10^7 M_{\sun}$ with a dilution factor of 3.5 (Fig.~\ref{lagen_mo}). This gives a $\chi^2$ = 8.41 for 13 degrees of freedom.  We obtain similarly good $\chi^2$ values for inclinations from $5-45^{\circ}$. With more data, we will be able to better measure the blue wing of the Fe~K lag, which will allow us to better constrain the disc inclination.  The Fe~K lag is sensitive to the height of the coronal source, and fits with larger heights give significantly worse results.  We note that the source height is strictly not a free parameter, as we only fit models with source heights of $h=2$, 5, 10, 20 and 40 $r_{\mathrm{g}}$. As we cannot set a confidence limit on the source height from lag fitting, we cannot statistically compare with the height found from spectral fitting, but we note that the two are roughly consistent, and both suggest a small source height. 

The Fe~K lag fit gives an estimate of a black hole mass of $1.3 \times 10^{7} M_{\sun}$, consistent to previous independent measurements.  The fitted disc inclination and source height are roughly the same as those derived from fitting the mean spectrum in Section~\ref{spec_sec}.  The dilution factor of 3.5 is high, though it is quite possible given that the power law in the mean energy spectrum likely contributes at frequencies outside of the frequency range of the Fe~K lag.

\section{Discussion and Conclusions}
\label{discuss}

\begin{figure}
\begin{center}
\includegraphics[width=7.5cm]{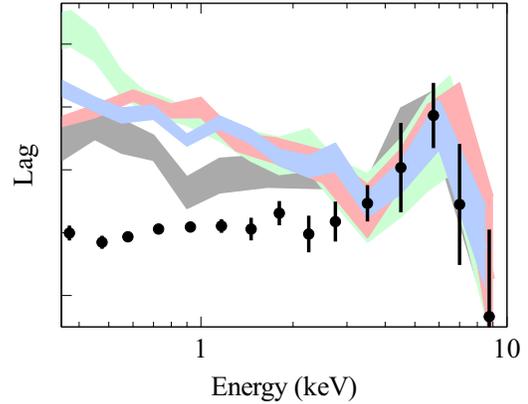}
\caption{The lag-energy spectrum of PG~1244+026 in black points, over-plotted with other lag-energy spectra: 1H0707-495 (blue), IRAS~13224-3809 (red), Ark~564 (green) and Mrk~335 (grey).  The masses (and amplitude of the Fe K lag) for these objects are all different, so the y-axis has been scaled to make the Fe~K lags match.  In PG~1244+026, the soft excess does not lag behind the continuum, as seen in these other Seyferts.}
\label{lagen_stack}
\end{center}
\end{figure}

The discovery of the Fe~K lag presented here is a strong indication that there is relativistic reflection off a highly ionised accretion disc in this object.  X-ray reverberation gives the scale of the X-ray source and accretion disc in physical units (e.g. in kilometres rather than gravitational units), and so, if we understand the geometry of the system, we can use the amplitude of the lag as an indicator of the black hole mass \citep*[See Fig.~12 of ][, which shows the linear scaling relationship between Fe~K lag amplitude and black hole mass for the seven sources with known Fe~K lags]{kara13c}.  We find that for PG~1244+026, the amplitude of the Fe~K lag scales with black hole mass, as well, though we note that independent mass estimates can vary by an order of magnitude in this source.  We briefly outline other mass estimates of PG~1244+026, but direct the reader to Section~6.3 of \citet{jin13} for a more detailed discussion.  In that work, the mass obtained from fitting the spectral energy distribution is $M_{\mathrm{BH}}=1.6 \times 10^{7} M_{\sun}$.  Optical reverberation measurements put it at a lower black hole mass of $4.8 \times 10^6 M_{\sun}$ \citep{vestergaard06}. However, taking into account the effects of radiation pressure increase the mass estimate to $1.8 \times 10^7 M_{\sun}$ \citep{marconi08}.  Using the hard X-ray variance and black hole mass correlation suggests a black hole mass of $0.5-1.5 \times 10^7 M_{\sun}$ \citep{ponti12}.  Our General Relativistic ray-tracing fits find a mass of $1.3\times 10^{7} M_{\sun}$, which is at the high end of previous estimates, though is entirely consistent.

%Given this lag amplitude and black hole mass, the X-ray source is at a height of $\sim 5-6 r_{\mathrm{g}}$.  Our General Relativistic ray-tracing modelling confirms that the lag requires a source height of $5~r_{\mathrm{g}}$.

We note that in most sources, we see a hard lag at low frequencies, which we cannot constrain here with the data available.  The hard lag shows a largely featureless lag-energy spectrum (see Fig.~4 in Kara et al, 2013c for an example in Ark~564.)  This lag is interpreted as being due to fluctuations in the disc accretion rate.  Fluctuations propagate inwards on the viscous timescale, and get transferred up to the corona (presumably by magnetic fields).  This causes the coronal emission at larger radii to respond before the coronal emission at smaller radii \citep{kotov01,arevalo06}.  If the harder emission is produced at the smaller radii (i.e. because of higher electron temperature closer to the core), then the hard emission will lag behind the soft emission produced further out.  It is clear from the lag-energy spectrum in Fig.~\ref{totlagen} that the lag is not due to the propagation lag because of the sharp drop above 6.5~keV, indicative of a strong broad Fe~K reflection feature at these frequencies. Furthermore, If we were probing a propagation lag that was being diluted by some smaller scale reflection, then the reflection component would have to be nonphysically large to cancel out the large amplitude propagation lag. This diluting reflection lag would have to be very narrow, and we would expect to see dilution effects at lower energies (from the red wing of the line) as well.  All this strongly supports a reverberation interpretation for the lags at these frequencies.

The Fe~K lag in PG~1244+026 also shows frequency dependence.  At lower frequencies, where the timescale of variability indicates a larger emitting region, the longest lag is at the line centroid of the Fe~K lag. This is expected as the larger emitting region irradiates larger radii of the accretion disc. At higher frequencies, from a more compact emitting region, the peak of the lag is at $\sim 4-5$~keV, the red wing of the line.  The compact emitting region will irradiate the smallest radii, where the gravitational redshift will be greatest.  The amplitude of the lag is smaller, as the light travel time is less. 

Fe~K lags have been seen in eight Seyfert galaxies thus far, and we over-plot four of those in the contours of Fig.~\ref{lagen_stack} \citep[as in Fig.~11 of ][]{kara13c}. We over-plot in black points the lag-energy spectrum of PG~1244+026 for comparison.  The shape of the Fe~K lag is similar, but the soft energy lags have a very different behaviour. The soft excess does not lag behind the continuum as seen in the other four sources.

The lag-energy spectrum provides us with an additional dimension---time---with which to interpret the mean spectrum.  PG~1244+026 has a strong soft excess above the continuum, but, as noted, that soft excess shows no time delay with respect to the continuum. Naively, we expect that if the soft excess was predominantly caused by broadened emission lines from the irradiated accretion disc, then there should be time delays corresponding to the light-travel time from the X-ray source.  PG~1244+026 shows a clear broad Fe~K lag, a strong signature of relativistic reflection, which is, for some reason, not seen at soft energies. There are several possibilities for this lack of soft lag. It could be that the accretion disc in this source has a lower ionisation parameter, which means that most of the photons are not absorbed, and therefore we only see emission from the strongest line, the Fe~K line.  

Another possibility, which we have modelled in Fig.~\ref{spec}, is that the soft excess is dominated by emission from a jet. In this scenario, we see the synchrotron tail of the jet, which we model phenomonologically as a steep power law.  The corona (or base of the jet) is responsible for irradiating the accretion disc.  Therefore, some of the jetted emission is correlated with the reflected emission, while some of it is correlated only with itself.  This explains the result by \citet{jin13}, that the 4--10~keV band is not strongly correlated with the soft excess at frequencies above $2 \times 10^{-4}$~Hz, where reflection contributes more. 
If the majority of the soft excess is caused by emission from the jet, then the majority of the soft excess at this frequency has zero lag with respect to the full reference band. There is some non-zero lag caused by reflection in the soft excess, but it is diluted by this dominating jetted, zero-lag component.

At high energies, however, where the bulk of the continuum comes from the base of the jet, which has a harder spectrum, the accretion disc is irradiated, and we measure the Fe~K lag.  Statistically, this model gives a good fit to the time-integrated energy spectrum, and could be possible as this source has some radio emission \citep{rafter09}.  There is much work to be done in understanding the origin of the soft excess, and we offer this model for PG~1244+026 as a possible explanation that is consistent with the spectral-timing results.

We emphasise that all of this work has been done with one orbit of {\em XMM-Newton} data.  Similar analyses in other objects have required much more data (i.e. 500~ks in Ark 564, 1.3~Ms in 1H0707-495).  PG~1244+026 is a remarkable source with a high count rate, and its clear Fe~K lag signature allows us to probe the relativistic effects close to the central black hole.  

\section*{Additional Note}
After submission, we discovered that X-ray time lags in PG~1244+026 were simultaneously studied by \citet{alston13}.  That submitted paper focuses on the lags associated with the soft excess, whereas this paper focuses on the Fe~K reverberation lag.  The results in these two papers are consistent, with the only difference being in the choice of reference band.  In this paper, we use a large reference band from 0.3--10~keV, so as to maximize the signal-to-noise at the Fe~K band, whereas in \citet{alston13}, they choose the 1.2--4~keV band, in order to track changes in the soft excess with frequency.  The results from these papers show that both relativistic reflection and propagation of mass accretion rate fluctuations are important effects in understanding the time lags.  These papers also show the complexities of the soft excess, and the importance of looking at the Fe~K band for probing strong gravity.  

%\section*{Acknowledgements}

%This work is based on observations obtained with {\em XMM-Newton}, an ESA science mission with instruments and contributions directly funded by ESA Member States and NASA.  EK thanks the Gates Cambridge Scholarship. ACF thanks the Royal Society.

%\bsp

\label{lastpage}


\begin{thebibliography}{99}


\bibitem[\protect\citeauthoryear{Alston, Done \& Vaughan}{2013}]{alston13} 
Alston W., Done C., Vaughan S., MNRAS submitted, arXiv:submit/0851673 

\bibitem[\protect\citeauthoryear{Ar{\'e}valo
\& Uttley}{2006}]{arevalo06} Ar{\'e}valo P., Uttley P., 2006, MNRAS, 367, 801


\bibitem[\protect\citeauthoryear{Cackett et 
al.}{2013}]{cackett13} Cackett E.~M., Zoghbi A., Reynolds C., 
Fabian A.~C., Kara E., Uttley P., Wilkins D.~R., 2013, arXiv, 
arXiv:1311.2997 


\bibitem[\protect\citeauthoryear{Crummy et al.}{2006}]{crummy06} 
Crummy J., Fabian A.~C., Gallo L., Ross R.~R., 2006, MNRAS, 365, 1067 


\bibitem[\protect\citeauthoryear{Fabian et al.}{1989}]{fabian89} 
Fabian A.~C., Rees M.~J., Stella L., White N.~E., 1989, MNRAS, 238, 729 

\bibitem[\protect\citeauthoryear{Fabian et al.}{2013}]{fabian13} 
Fabian A.~C., et al., 2013, MNRAS submitted 

\bibitem[\protect\citeauthoryear{Guilbert 
\& Rees}{1988}]{guilbert88} Guilbert P.~W., Rees M.~J., 1988, MNRAS, 233, 475 

\bibitem[\protect\citeauthoryear{Dauser et al.}{2013}]{dauser13} 
Dauser T., Garcia J., Wilms J., B{\"o}ck M., Brenneman L.~W., Falanga M., 
Fukumura K., Reynolds C.~S., 2013, MNRAS, 430, 1694 

\bibitem[\protect\citeauthoryear{Jansen et al.}{2001}]{jansen01}
Jansen F., et al., 2001, A\&A, 365, L1



\bibitem[\protect\citeauthoryear{Jin et al.}{2012}]{jin12} 
Jin C., Ward M., Done C., Gelbord J., 2012, MNRAS, 420, 1825 



\bibitem[\protect\citeauthoryear{Jin et al.}{2013}]{jin13} 
Jin C., Done C., Middleton M., Ward M., 2013, arXiv, arXiv:1309.5875 


\bibitem[\protect\citeauthoryear{Kara et al.}{2013a}]{kara13a} 
Kara E., Fabian A.~C., Cackett E.~M., Uttley P., Wilkins D.~R., Zoghbi A., 
2013, MNRAS, 434, 1129 


\bibitem[\protect\citeauthoryear{Kara et al.}{2013b}]{kara13b} 
Kara E., Fabian A.~C., Cackett E.~M., Miniutti G., Uttley P., 2013, MNRAS, 
430, 1408 


\bibitem[\protect\citeauthoryear{Kara et al.}{2013c}]{kara13c} 
Kara E., Fabian A.~C., Cackett E.~M., Steiner J.~F., Uttley P., Wilkins 
D.~R., Zoghbi A., 2013, MNRAS, 428, 2795 

\bibitem[\protect\citeauthoryear{Kotov, Churazov,
\& Gilfanov}{2001}]{kotov01} Kotov O., Churazov E., Gilfanov M., 2001, MNRAS, 327, 799


\bibitem[\protect\citeauthoryear{Lightman 
\& White}{1988}]{lightman88} Lightman A.~P., White T.~R., 1988, ApJ, 335, 57 


\bibitem[\protect\citeauthoryear{Marconi et 
al.}{2008}]{marconi08} Marconi A., Axon D.~J., Maiolino R., Nagao 
T., Pastorini G., Pietrini P., Robinson A., Torricelli G., 2008, ApJ, 678, 
693 


\bibitem[\protect\citeauthoryear{Nowak et al.}{1999}]{nowak99}
Nowak M.~A., Vaughan B.~A., Wilms J., Dove J.~B., Begelman M.~C., 1999,
ApJ, 510, 874



\bibitem[\protect\citeauthoryear{Ponti et 
al.}{2012}]{ponti12} Ponti G., Papadakis I., Bianchi S., Guainazzi M., Matt G., Uttley P., Bonilla N.~F., 2012, A\&A, 542, A83 

\bibitem[\protect\citeauthoryear{Porquet et 
al.}{2004}]{porquet04} Porquet D., Reeves J.~N., O'Brien P., Brinkmann W., 2004, A\&A, 422, 85 



\bibitem[\protect\citeauthoryear{Rafter, Crenshaw, 
\& Wiita}{2009}]{rafter09} Rafter S.~E., Crenshaw D.~M., Wiita P.~J., 2009, AJ, 137, 42 


\bibitem[\protect\citeauthoryear{Reynolds et 
al.}{1999}]{reynolds99} Reynolds C.~S., Young A.~J., Begelman 
M.~C., Fabian A.~C., 1999, ApJ, 514, 164 

\bibitem[\protect\citeauthoryear{Ross 
\& Fabian}{2005}]{ross05} Ross R.~R., Fabian A.~C., 2005, MNRAS, 358, 211 


\bibitem[\protect\citeauthoryear{Str{\"u}der et
al.}{2001}]{struder01} Str{\"u}der L., et al., 2001, A\&A, 365, L18


\bibitem[\protect\citeauthoryear{Vestergaard 
\& Peterson}{2006}]{vestergaard06} Vestergaard M., Peterson B.~M., 2006, ApJ, 641, 689 


\bibitem[\protect\citeauthoryear{Wilkins 
\& Fabian}{2013}]{wilkins13} Wilkins D.~R., Fabian A.~C., 2013, MNRAS, 430, 247 



\bibitem[\protect\citeauthoryear{Zoghbi, Uttley, 
\& Fabian}{2011}]{zoghbi11} Zoghbi A., Uttley P., Fabian A.~C., 2011, MNRAS, 412, 59 


\bibitem[\protect\citeauthoryear{Zoghbi et al.}{2012}]{zoghbi12}
Zoghbi A., Fabian A.~C., Reynolds C.~S., Cackett E.~M., 2012, MNRAS, 422,
129


\bibitem[\protect\citeauthoryear{Zoghbi et al.}{2013}]{zoghbi13} 
Zoghbi A., Reynolds C., Cackett E.~M., Miniutti G., Kara E., Fabian A.~C., 
2013, ApJ, 767, 121 



\end{thebibliography}
\end{document}